\documentclass[conference]{IEEEtran}
\IEEEoverridecommandlockouts
\usepackage{cite}
\usepackage{amsmath,amssymb,amsfonts}
\usepackage{algorithmicx}
\usepackage{algorithm}
\usepackage{algpseudocode}  
\usepackage{amsthm}
\usepackage{subfigure}
\usepackage{graphicx}
\usepackage{textcomp}
\usepackage{amsfonts,amssymb}
\usepackage{xcolor}
\def\BibTeX{{\rm B\kern-.05em{\sc i\kern-.025em b}\kern-.08em
    T\kern-.1667em\lower.7ex\hbox{E}\kern-.125emX}}
\begin{document}
\bibliographystyle{IEEEtran}
\title{Edge-assisted Viewport Adaptive Scheme for real-time Omnidirectional Video transmission}

\author{\IEEEauthorblockN{Tao Guo,Xikang Jiang,Bin Xiang,Lin Zhang}
\\
\IEEEauthorblockA{\textit{School of Information and Communication Engineering} \\
\textit{Beijing University Posts and Telecommunications,Beijing,China}\\
\{guotao0706,jiangxikang,bingo6,zhanglin\}@bupt.edu.cn}
}

\maketitle

\begin{abstract}
Omnidirectional applications are immersive and highly interactive, which can improve the efficiency of remote collaborative work among factory workers. The transmission of omnidirectional video (OV) is the most important step in implementing virtual remote collaboration. Compared with the ordinary video transmission, OV transmission requires more bandwidth, which is still a huge burden even under 5G networks. The tile-based scheme can reduce bandwidth consumption. However, it neither accurately obtain the field of view(FOV) area, nor difficult to support real-time OV streaming. In this paper, we propose an edge-assisted viewport adaptive scheme (EVAS-OV) to reduce bandwidth consumption during real-time OV transmission. First, EVAS-OV uses a Gated Recurrent Unit(GRU) model to predict users' viewport. Then, users were divided into multicast clusters thereby further reducing the consumption of computing resources. EVAS-OV reprojects OV frames to accurately obtain users' FOV area from pixel level and adopt a redundant strategy to reduce the impact of viewport prediction errors. All computing tasks were offloaded to edge servers to reduce the transmission delay and improve bandwidth utilization. Experimental results show that EVAS-OV can save more than 60\% of bandwidth compared with the non-viewport adaptive scheme. Compared to a two-layer scheme with viewport adaptive, EVAS-OV still saves 30\% of bandwidth. The main part of the scheme is shown in https://github.com/kotorimaster/EVAS-OV.
\end{abstract}


\section{Introduction}
With the development of the fifth generation of mobile communication technology (5G), video, Virtual Reality(VR) and Augmented Reality(AR) applications will occupy over 90\% of 5G data usage. As an important part of VR, omnidirectional applications attract users with an immersive experience and high user interaction. In industrial scenarios, omnidirectional applications can be widely used for remote collaborative work and employee training\cite{hormaza2019line}. Real-time transmission of OV is the most critical step in implementing virtual remote collaboration. Due to frame rate, bit depth, etc., the bandwidth required to transmit OV is much higher than that of ordinary video with the same quality. In addition, when multiple users use head-mounted displays (HMD) to watch OV at the same time, bandwidth requirements will grow exponentially, and it will be challenging to meet bandwidth requirements even in 5G network.

In fact, the viewing angle of human eyes is about 120 degrees in horizontal and 90 degrees in vertical. When watching OV, the FOV area can only cover 1/6 of the entire video. Therefore, high-quality full-view OV transmission will cause a considerable waste of bandwidth. Researchers have proposed many tile-based transmission schemes \cite{petrangeli2017http}\cite{graf2017towards} based on user's viewports for this situation. However, these schemes need to divide OV into tiles by fixed duration and fixed size in advance, which is difficult to support real-time OV transmission. At the same time, this method doesn't take the impact of the projection method (Equirectangular Projection, a common projection method) on a user's FOV area into consideration. During OV transmission, all tiles within the FOV area need to be sent to the client. When a user's viewport deviates from the equator, the selection of viewport tiles becomes complicated, which we will explain in Section 3.

In a multi-user scenario, if all computing tasks are performed on the cloud server, it will be overburdened, offloading computing tasks to edge servers can reduce its computing burden. At the same time, most existing OV schemes are transmitted in unicast. When multiple users watch the same video, multiple transmissions of the same video will further increase the network burden. In fact, viewports are similar when users watch the same OV. By clustering user viewports, OV frames of the same cluster can be transmitted using the multicast mode, which can reduce the consumption of bandwidth resources by the repeated transmission of OV.

\begin{figure*}[!htbp]
\centerline{\includegraphics[width = 0.90\textwidth]{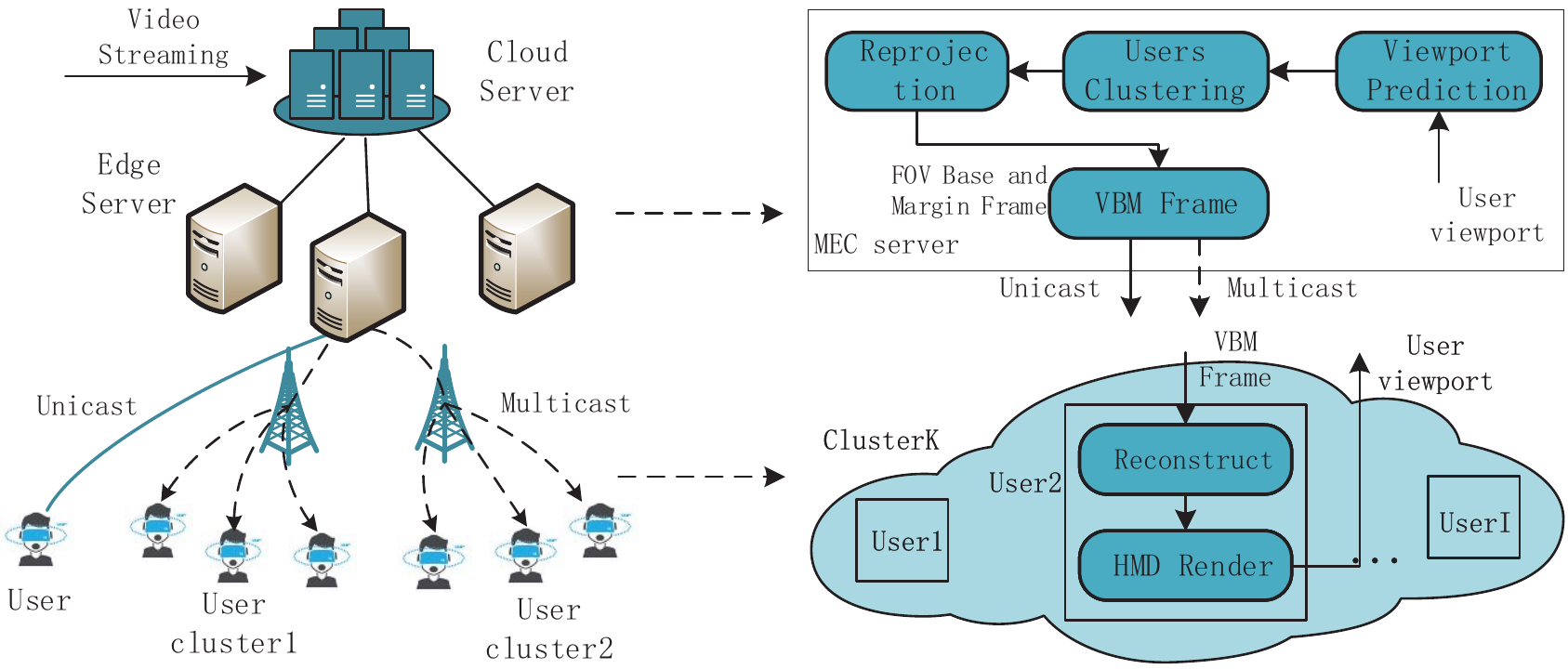}}
\caption{Network architecture and Processing details}
\label{fig}
\end{figure*}

In this paper, we propose a viewport-based and edge-assisted OV transmission scheme. The network architecture is shown in Figure 1. The cloud server is responsible for video content distribution, and the edge server is responsible for viewport prediction, user clustering, and generation of reprojection frame. Specifically, we deployed a viewport prediction module, a user clustering module, and a frame processing module on the edge server. A GRU \cite{chung2014empirical} neural network(NN) model is run on the prediction module to predict users' viewports. The user clustering module divides users in similar viewports into clusters, and OV frames of the same cluster can be transmitted using multicast mode. The frame processing module reprojects the OV frame according to the user viewport and generates combined frames. The specific details will be described in Section 3. We prototyped the scheme and evaluated its performance. The results show that it can save more than 60\% of bandwidth compared to non-viewport adaptive schemes. 

The contributions of this paper are summarized as follows:
\begin{itemize}
    \item We propose a real-time OV transmission optimization scheme based on edge computing. It can accurately obtain a user's FOV area from pixel level, saving bandwidth during transmission.
    \item We use a GRU network to predict user's viewport and propose a redundant strategy -- VBM(FOV, Base, Margin) frame to reduce the impact of prediction errors on users' FOV area.
    \item We use a clustering method to group users with similar viewports. OV frames of the same cluster can be transmitted using the multicast mode, which can effectively reduce the occupation of computing and bandwidth resources.
\end{itemize}

The rest of this paper is summarized as follows. Section 2 introduces some related work on viewport prediction, OV transmission, and multi-user clustering. Section 3 introduces our proposed scheme, including viewport prediction, user clustering, and VBM frame generation. In Section 4, we introduce the experimental configuration and give experimental results and our analysis. Section 5 summarizes this paper and gives conclusions.

\section{Related Work}

A viewport-based transmission scheme can effectively reduce the bandwidth occupation when transmitting OV. However, the viewport prediction error will cause the transmission of the FOV area that does not match the user's viewport to HMD  or show a blank FOV. Some machine learning methods are used for the users' viewport prediction. Bao Y et al. \cite{bao2016shooting} uses a linear regression (LR) model to fit the variation of a user's viewports and achieve good performance in 100-500ms. In paper \cite{aladagli2017predicting}, the author takes content characteristics such as user's preferences, occupation, gender, age, and other factors into consideration, and describes the relationship between future viewports and historical viewports as a non-linear and long-term dependency, and uses a saliency algorithm to predict viewport.  

Transmitting the high-quality and full-view OV is challenging. Most optimization schemes focus on tile-based transmission schemes. S.Petrangeli al. \cite{petrangeli2017http} adopts a full-pass advanced scheme, viewport tiles have the highest possible bitrate while giving other tiles a lower but not the lowest bit rate. In \cite{graf2017towards}, viewports, adjacent regions, and outer regions are defined, and the available bandwidth is allocated to tiles in different regions based on regional priorities. Sun L et al. \cite{sun2019two} implements a system that can effectively adapt to network fluctuations and viewpoint prediction errors. In paper\cite{shi2019freedom}, researchers have proposed a method that neither uses a tile scheme nor relies on viewport prediction. They implemented a system that can display high-quality OV at 60 FPS on mobile devices.

When multiple users watch OV at the same time, the unicast transmission mode will cause multiple transmissions of the same video, wasting a lot of bandwidth. In paper\cite{long2018optimal} users' common tiles are transmitted in multicast, and other tiles are transmitted in unicast. Paper\cite{yuan2017ag} considers a method of combining multicast and transcoding based on the tile scheme. Server multicasts highest bitrate tile to users and users obtain required bitrate tile through client transcoding. Paper\cite{yang2019cmu} applies the clustering method to a multi-user bitrate adaptive system. We consider clustering multiple users based on user viewport similarity. Multicast transmission mode after clustering can effectively reduce the occupation of bandwidth and computing resources during transmission.

\section{Proposed Scheme}

\subsection{Scheme Overview}

We propose a pixel-level transmission optimization scheme to reduce bandwidth consumption during OV transmission. Because users' viewports prediction and generation of VBM frames are high-computation processes, which will cause huge computing pressure on the cloud server, we offload these computing tasks to edge servers. The specific processing details are shown in fig 1. 

Edge server mainly performs viewport prediction, user clustering, and VBM frame generation. First, we use a GRU neural network to predict users' viewports(\S B). The clustering module divides users with similar viewports into a cluster and selects the average value of the user's viewport positions in the cluster as the center of the cluster. Users in one cluster receive frames in the mode of multicast, and individual users receive frames with unicast mode(\S C). Then, the frame processing module re-projects OV frames, to accurately extract the FOV frame. The margin and base frame will be combined with the FOV frame to generate a VBM frame for the convenience of encoding and transmission(\S D).

\subsection{Viewport Prediction}\label{AA}
Good user experience for VR applications requires motion-to-photon delay 15 to 20 ms, which is difficult to be satisfied in most transmission schemes. Prefetching user's viewport through viewport prediction can effectively solve this problem. Some schemes use LR models for prediction and obtain good results in a short prediction time. However, as mentioned in section 2, the relationship between future viewports and historical viewports is non-linear and long-term dependency. When the time for prediction increased from 0.5s to 1s, accuracy will be greatly reduced.

In our scheme, we use the GRU neural network for viewport prediction. GRU has two gates (update gate, reset gate). The update gate is used to control the degree to which the state information of the previous moments enters into the current state. The reset gate is used to control the degree of ignoring status information of the previous moments. These two gates can handle long-term dependency problems between data, so GRU can obtain good results in continuous data prediction. We use (yaw, roll, pitch) to represent user's viewport ${{V_{{\rm{t}}}}}$. Since HMD uses unit quaternion data(q0,q1,q2,q3) as output information, we need to convert it to (yaw, pitch, roll). Bao Y et al. \cite{bao2017motion} pointed out that the prediction of each angle can be independent of others, so we make predictions for each of them separately. 

Our predicting model consists of a hidden layer and two GRU layers. User's historical viewport in the past L time intervals can be expressed as $\left( {{V_{\rm{t}}}{\rm{,}}{{\rm{V}}_{{\rm{t - 1}}}}{\rm{,}}...{\rm{,}}{{\rm{V}}_{{\rm{t - L + 1}}}}} \right)$, ${V_t}$ is user's viewport at time t, and predicted viewport can be expressed as ${{V_{{\rm{t + T}}}}}$, where T is the predicted time interval.

Generally, viewport moves more frequently in horizontal when a user watches the OV. As a result, the yaw prediction will be more difficult than pitch and roll. At the same time, viewport prediction error may cause a FOV area that does not match the user's actual FOV to appear on HMD. When a user's viewport change, fluctuation of image quality in the FOV area can easily make users feel dizzy and affect their experience. To guarantee users' experience, we adopt a redundant strategy to ensure that image quality in the user's viewport area won't be seriously degraded even if a prediction error occurs.  

\subsection{User Clustering}

Most existing transmission schemes use unicast for OV transmission. When multiple users watch OV at the same time, the unicast mode will cause a video to be transmitted multiple times on link, which occupies a lot of bandwidth resources. In fact, different user's viewports are similar when watching the same OV. Therefore, distributing OV streaming in the mode of multicast can reduce the bandwidth consumption and the occupation of computing resources at the edge server.

We use the Density-Based Spatial Clustering of Application with Noise (DBSCAN) for multi-user viewport clustering. It is a density-based clustering algorithm without specifying the number of clusters in advance. We choose the clustering radius to be 0.15 and the number of MinPts is set to 2. Before clustering, we need to transform viewport data into spatial coordinates for easy calculation.

\begin{figure}[!htbp]
\centerline{\includegraphics[width = 0.95\linewidth]{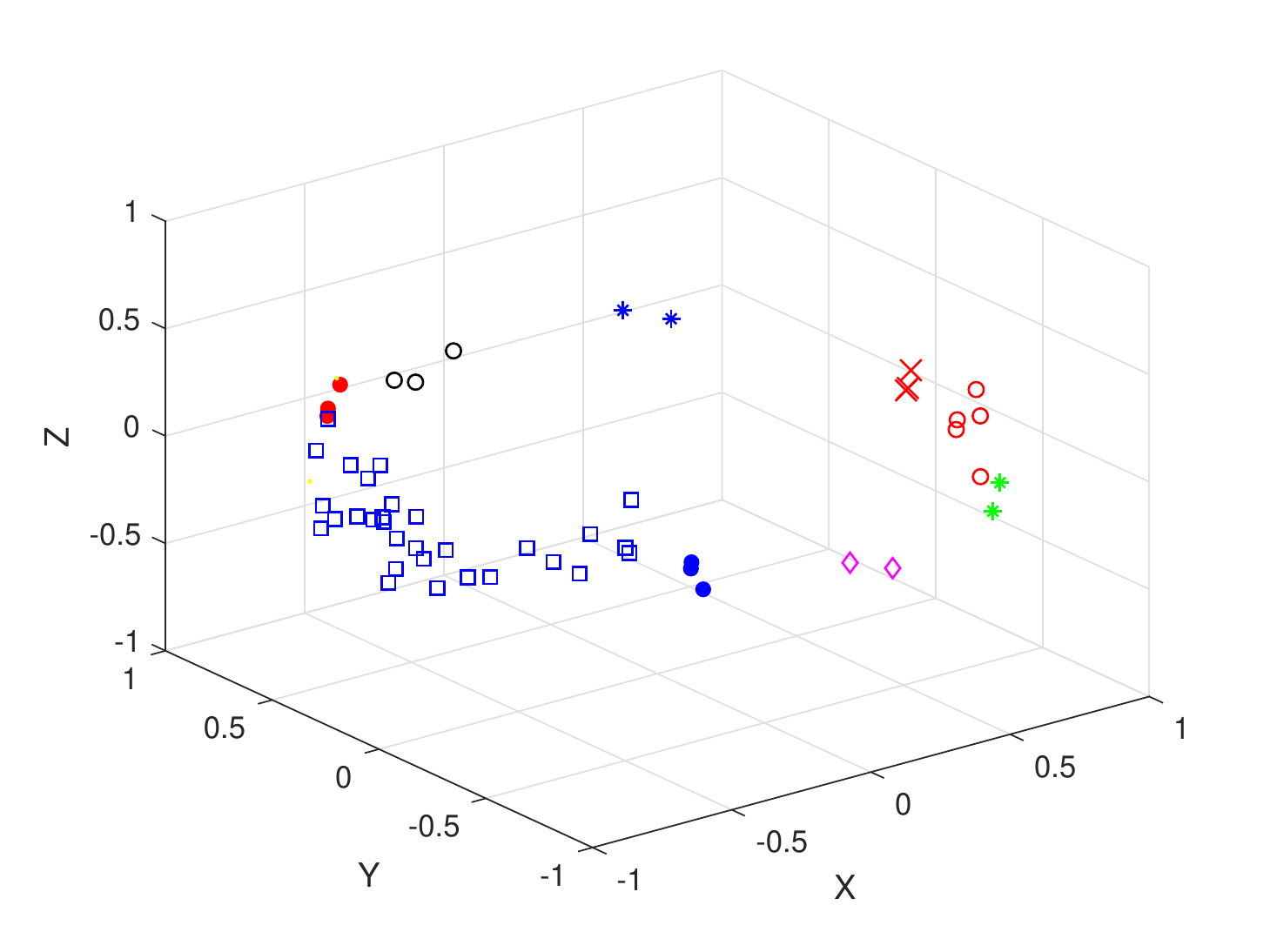}}
\caption{User clustering with $\omega = 0.8$}
\label{fig}
\end{figure}

When clustering users, we take users' head movement factor into consideration. If a user's head motion tends to move away from the cluster center, it shouldn't be involved in this cluster. We take the user's viewport position and head motion for joint clustering. Assuming the number of users is N, a user's viewport data can be expressed as $\left( {\mathop P\nolimits_{\rm{t}}^1 ,\mathop P\nolimits_{\rm{t}}^2 ,...,\mathop P\nolimits_{\rm{t}}^N } \right)$ , user's head motion data can be expressed as $\left( {\mathop V\nolimits_{\rm{t}}^1 ,\mathop V\nolimits_{\rm{t}}^2 ,...,\mathop V\nolimits_{\rm{t}}^N } \right)$.  

The distance matrix of two user's head movements is defined as:
\begin{equation}
M{V_{N \times N}}(i,j) = d(V_t^i - V_t^j) = {\left\| {v_t^i - v_t^j} \right\|_2},\forall i,j \in N
\end{equation}

And the distance matrix of two users' viewports positions is defined as:
\begin{equation}
M{P_{N \times N}}(i,j) = d(P_t^i - P_t^j) = {\left\| {P_t^i - P_t^j} \right\|_2},\forall i,j \in N
\end{equation}

We use a linear approach proposed in \cite{kan2019server} to combine two users' viewports positions and head movement distances. Distance parameter for clustering users can be expressed as:
\begin{equation}
\begin{array}{l}
{M_{N \times N}}(i,j) = \omega {\left[ {M{P_{N \times N}}(i,j)} \right]_n}\\
 + (1 - \omega ){\left[ {M{V_{N \times N}}(i,j)} \right]_n},\forall i,j \in N
\end{array}
\end{equation}
where ${\left[  \bullet  \right]_{\rm{n}}}$ is the normalization operator and $\omega$ is the weight parameter between head movement and the viewport position in users clustering. 

After clustering, users can be divided into $\left( {\mathop C\nolimits_1 ,\mathop C\nolimits_2 ,...,\mathop C\nolimits_K } \right)$, where K is the number of clusters. In Fig 2, 50 users are divided into 9 clusters. The center of ${\mathop C\nolimits_i}$ is mean of all viewports position within the cluster. Clustered users use multicast to obtain OV frames. Although this will have a negative impact on the FOV edge quality of clustered users, considering the gain brought by multicast, it is acceptable.

\subsection{VBM Frame Generation}

Common projection methods include Equirectangular Projection(ERP), Cubemaps and Equi-Angular Cubemap(EAC). ERP is currently the most commonly used method. By using the same number of sampling points to save data on each parallel, a rectangular video on a corresponding two-dimensional plane is obtained. Generally, we use (u, v) to represent the position of a pixel in the 2D plane, and use (x, y, z) to represent the position of a pixel on the sphere. The transformations between coordinates are shown as follows:

\begin{equation}
\left\{ \begin{array}{l}
x = \cos u * \sin v\\
y = \sin u * \sin v\\
z = \cos v
\end{array} \right.\
\end{equation}

\begin{equation}
\left\{ \begin{array}{l}
u = \frac{{[\arctan (\frac{y}{x} + \pi ) * {\rm{width}}]}}{{2*\pi }}\\
v = \frac{{[\arccos (z) * height]}}{\pi }
\end{array} \right.\
\end{equation}

However, ERP uses the same number of sampling points on each parallel line of a spherical video, the closer viewport to the polar region, the more redundant sampling points. This results in higher pixel density in the equatorial region and lower pixel density in the polar region when projected on a 2D plane. The difference in pixel density will have a huge impact on the acquisition of a user's FOV area. Figure 3(a) shows mapping range of user's FOV area on the 2D plane when user's pitch = 0°, 30°, and 90°, yaw and roll are both 0°. When the center of a user's viewport is at the equator, the shape of the FOV area is rectangular. As the viewport moves towards the pole area, the FOV area will stretch continuously. The above situation only considers changes in pitch. When a user uses an HMD to watch an OV, his head rotation will cause yaw, roll, and pitch to change at the same time, which will lead to more complex FOV areas. It is difficult to accurately obtain a user's FOV area using the tile-based transmission method.

\begin{figure}[!htbp] 
	\centering  
	\vspace{-0.35cm} 
	\subfigtopskip=5pt 
	\subfigbottomskip=10pt 
	\subfigcapskip=5pt 
	\subfigure[FOV area in different pitch]{
		\label{level.sub.1}
		\includegraphics[width=1\linewidth]{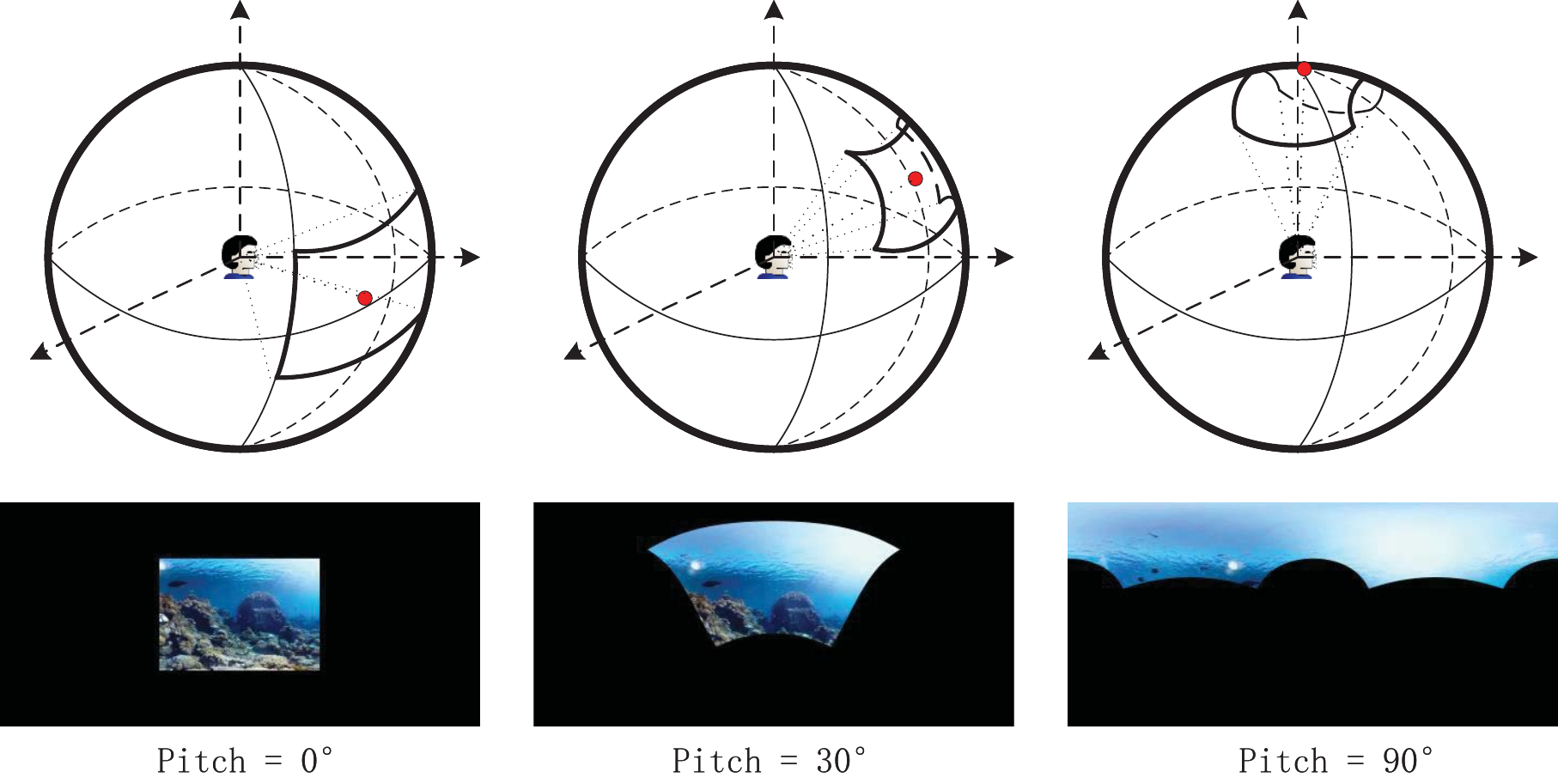}}
		
	\subfigure[Generation of VBM frame]{
		\label{level.sub.2}
		\includegraphics[width=1\linewidth]{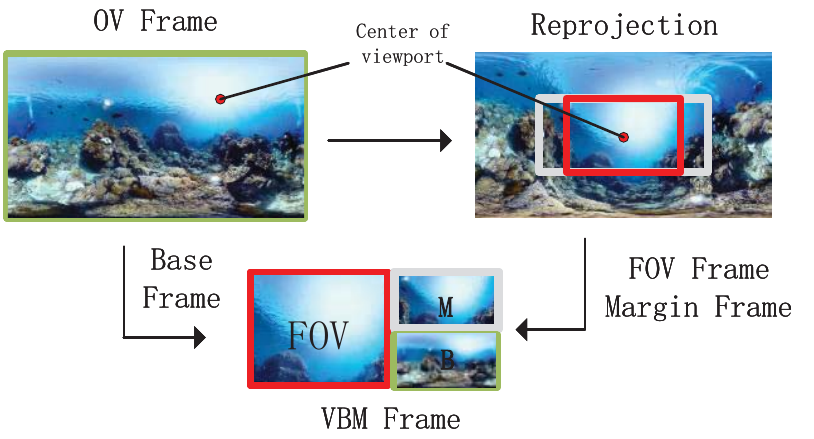}}

	\caption{FOV area and Combined VBM frame}
	\label{level}
\end{figure}

Extracting irregular FOV part is difficult. To simplify the extraction of the FOV part, we re-project the OV frame and remap the user's viewport to the center of the omnidirectional frame. The mapping relationship is:

\begin{equation}
\psi' = W \psi 
\end{equation}

in which $\psi'$ is the new rectangular coordinate after reprojection, $\psi$ is original rectangular coordinate. W is the reprojection matrix generated based on parameters of a user's head rotation(yaw, roll, pitch). First, we remap the OV frame to a sphere, then use the reprojection matrix to rotate the sphere and generate a new rectangular projection. After reprojection, the center of the user's viewport is at the center of the new omnidirectional frame. The user's FOV area can be easily obtained.

A viewport-based transmission scheme can effectively reduce bandwidth consumption during OV transmission. However, this also means that when a viewport prediction error occurs, frames that do not match the actual FOV can severely impact the users' experience. Therefore, we adopted a redundant strategy to reduce the impact of viewport prediction errors on users' experience.

We divide the omnidirectional frame into three parts: base, margin, and FOV. The selection of FOV size is related to HMD. Here we select the FOV area as a 120°×90° area. First, after the OV frame reprojection, the FOV area will be located in the center of the rectangular plane, and we can easily obtain the user's FOV. The FOV area can be cropped directly without any quality loss. Generally, areas outside the FOV don't appear in the user's viewport, so we can reduce its quality as much as possible. We deploy downsampling for the entire omnidirectional frame to generate a low-quality base frame (B-frame), the downsampling rate is 1/4. B-frame is used to ensure the lowest image quality of FOV under any head movement. We use a margin frame to guarantee users' quality of experience(QoE). Based on the analysis of viewport predictions, the range of margin area is set to $ \pm {30^ \circ }$ in the yaw direction. After downsampling, the size of the margin frame is the same as the base frame and the height is half of the FOV frame. The quality of the margin frame is lower than the FOV frame but still higher than the base frame. As shown in fig3(b), we formulate a combination rule and generated a VBM frame to facilitate encoding and transmission. After receiving the VBM frame, the client will reconstruct it and generate a reconstructed OV frame for rendering.

\section{Evaluation and Performance Results}
We verified the feasibility of EVAS-OV and compared it with other schemes. 

\subsection{Experiment Setup}
We use an open-source dataset\cite{corbillon2017360} to train the viewport prediction model on a server configured with 6 GPUs. It includes head movement data when 59 users watch OV, and there are 600 log files in total. The edge server used in this scheme is Dell T630, its operating system is Ubuntu 16.04. The GPU model on the edge server is NVIDIA GeForce GTX 1080Ti. On the display side, we use an Oculus Rift as the head-mounted device.

\begin{table}[h]  
  \centering  
  \caption{Video Information}  
    \begin{tabular}{|c|c|c|c|c|}  
    \hline  
    No.	& Content	& duration & Size(MB) & FPS 	\\
    \hline  
    1&Ocean&06:52&963&30\cr\hline 
    2&Roller Coaster&03:26&396&30\cr\hline
    3&Street&01:31&169&30\cr\hline
    4&Grassland1&02:49&250&30\cr\hline
    5&Grassland2&01:41&75.8&30\cr
    \hline  
    \end{tabular}  
\end{table}

In comparative experiments, We select OVs in five different scenarios from the dataset. Scenarios include street, roller coaster, grassland, and ocean. The video information is given in Table 1. All video's resolution is 3840 * 2048. We take 20 seconds clips from each of five videos, and the frame rate is 30 FPS. We use FFmpeg as our video processing tool and H.264/AVC as our encoding standard.

\subsection{Evaluation results}

\subsubsection{Viewport prediction}
We make an independent prediction on yaw, roll, and pitch. Since a user's head movement is random, it is difficult to predict viewport for a long time, we choose prediction interval length T = 1s. In fig 4(a), We compare prediction results of the GRU neural network with the LR model. The result shows that the prediction accuracy of the GRU model in Yaw reaches 90\%, which is 7.5\% higher than the LR model. As shown in fig 4(b), since a user's head rotation occurrence less in pitch and roll direction, the accuracy of pitch and roll prediction is significantly higher than yaw. The prediction accuracy of the roll is close to 99\%, while the prediction accuracy of pitch also reached 98\%.

\begin{figure}[!htbp] 
	\centering  
	\vspace{-0.35cm} 
	\subfigtopskip=5pt 
	\subfigbottomskip=10pt 
	\subfigcapskip=-5pt 
    \subfigure[prediction error CDF]{
		\label{level.sub.1}
		\includegraphics[width=0.45\linewidth]{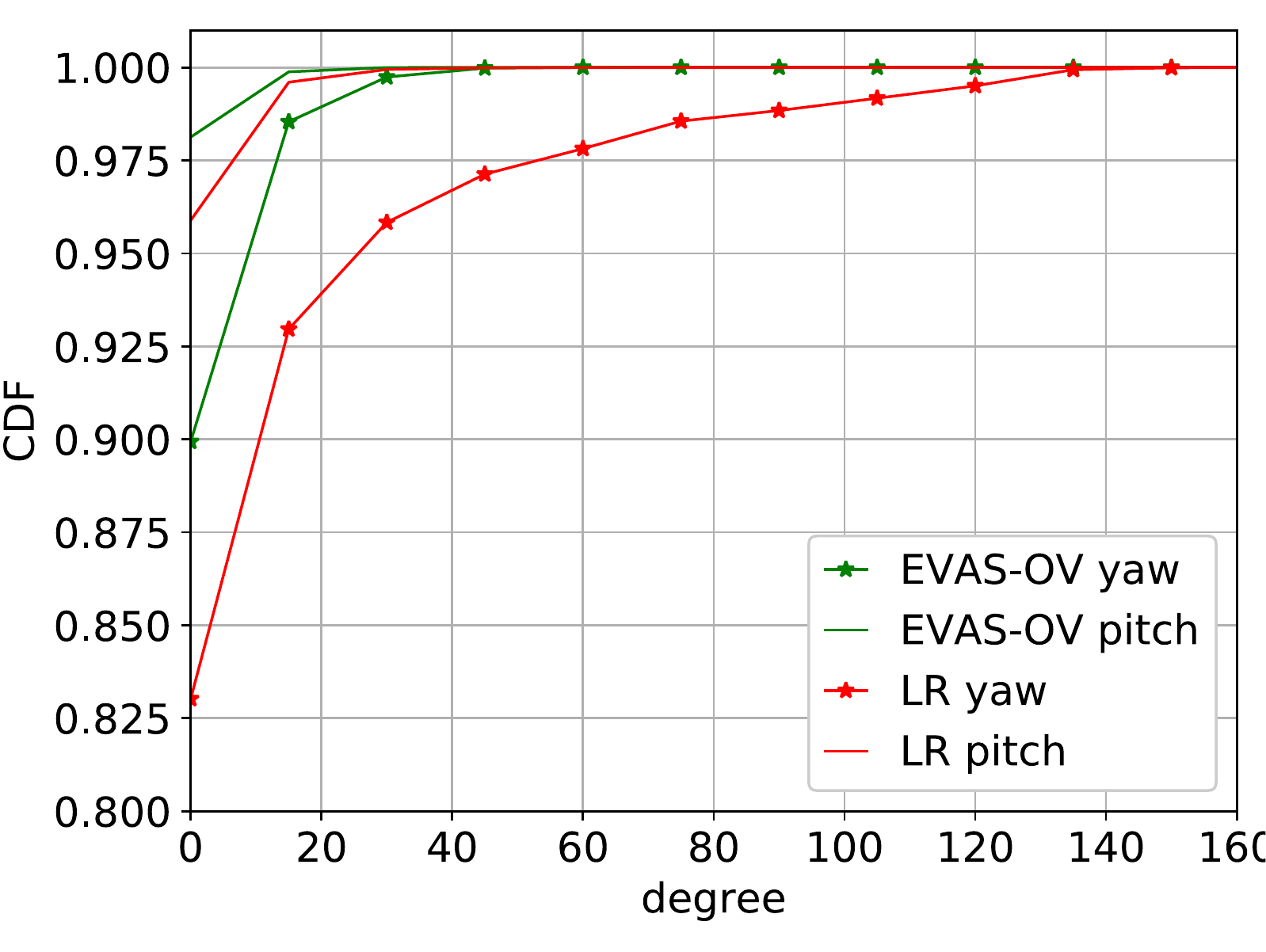}}
	\quad 
	\subfigure[prediction in yaw,roll,pitch]{
		\label{level.sub.2}
		\includegraphics[width=0.45\linewidth]{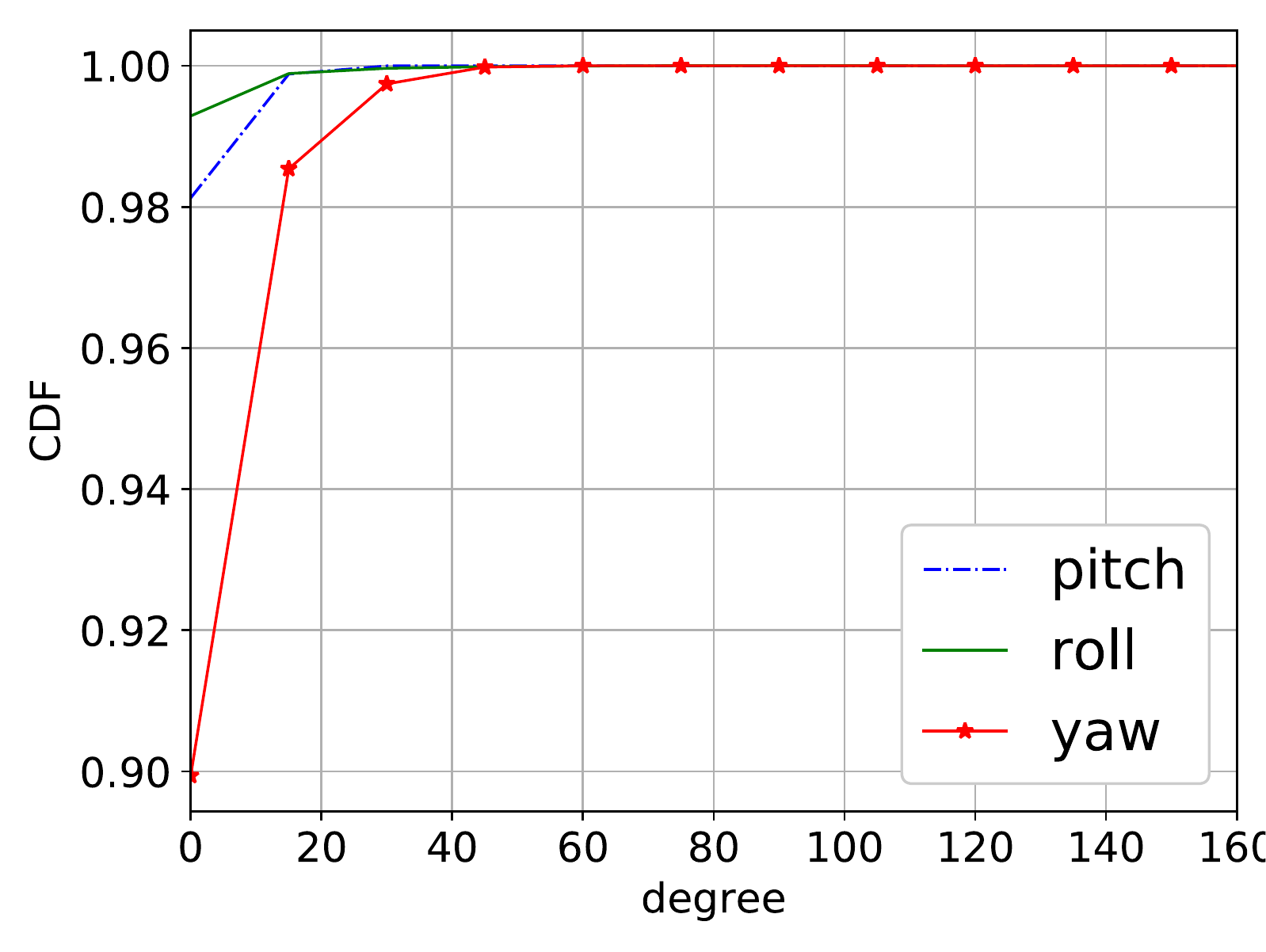}}
	\quad
	\caption{viewport prediction results}
	\label{level}
\end{figure}

\begin{figure}[!htbp] 
	\centering  
	\vspace{-0.35cm} 
	\subfigtopskip=5pt 
	\subfigbottomskip=10pt 
	\subfigcapskip=-5pt 
	\subfigure[FOV quality loss MSE with different Videos]{
		\label{level.sub.3}
		\includegraphics[width=0.45\linewidth]{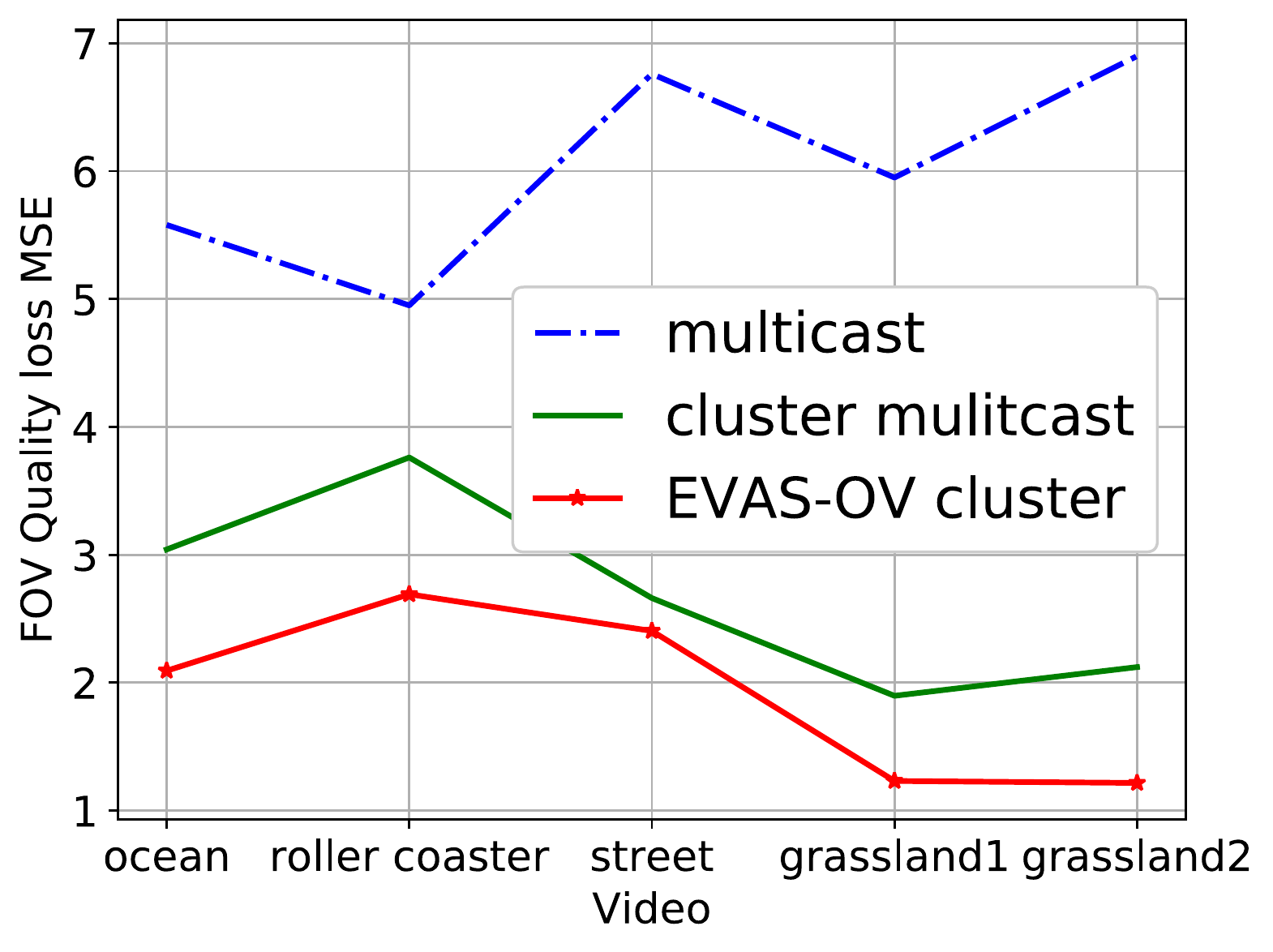}}
	\quad
	\subfigure[FOV quality loss MSE with different number of users]{
		\label{level.sub.4}
		\includegraphics[width=0.45\linewidth]{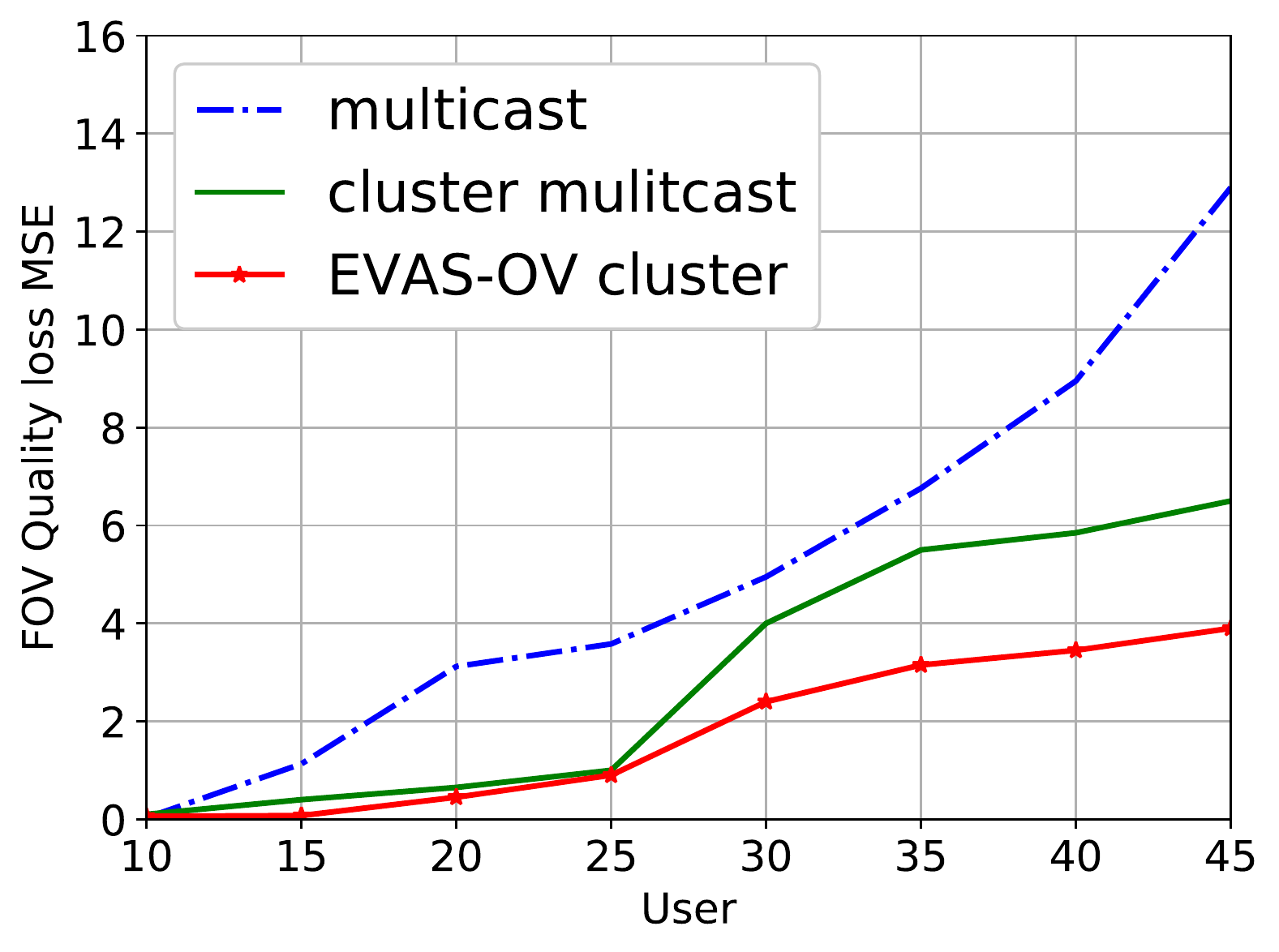}}

	\caption{Average FOV quality loss}
	\label{level}
\end{figure}

\subsubsection{User clustering}
Figure 5 shows the effect of clustering on user FOV under different numbers of users and different OVs. We use a multicast method and clustered multicast as the comparison method. The multicast method means directly adopts the multicast mode for frame transmission without clustering. The cluster multicast method only considers user viewports to cluster users. EVAS-OV performs joint clustering based on user viewport and head movement, here we select the weight index $\omega = 0.8$. Fig5(a) shows the FOV quality loss at different OVs. Our clustering method has less impact on users' FOV qualities. Fig5(b) shows the FOV quality loss when clustering different numbers of users. When the number of users exceeds 25, our proposed method makes users' FOV quality significantly improved.

\subsubsection{Bandwidth savings}
To verify the efficiency of our scheme, we compare it with a non-viewport scheme and a two-layer transmission scheme: 
\begin{itemize}
\item Non-viewport adaptive scheme: a current widely used transmission scheme in omnidirectional applications, sending full view OV to the client.
\item Two-layer scheme\cite{duanmu2017view}: The base layer is the full-view OV with the lowest quality, and the enhancement layer is used to improve viewport quality. The quality of the base layer is equal to the base frame in our scheme.
\end{itemize}

\begin{figure}[!htbp] 
	\centering  
	\vspace{-0.35cm} 
	\subfigtopskip=5pt 
	\subfigbottomskip=10pt 
	\subfigcapskip=-5pt 
    \subfigure[Bandwidth Savings]{
		\label{level.sub.1}
		\includegraphics[width=0.45\linewidth]{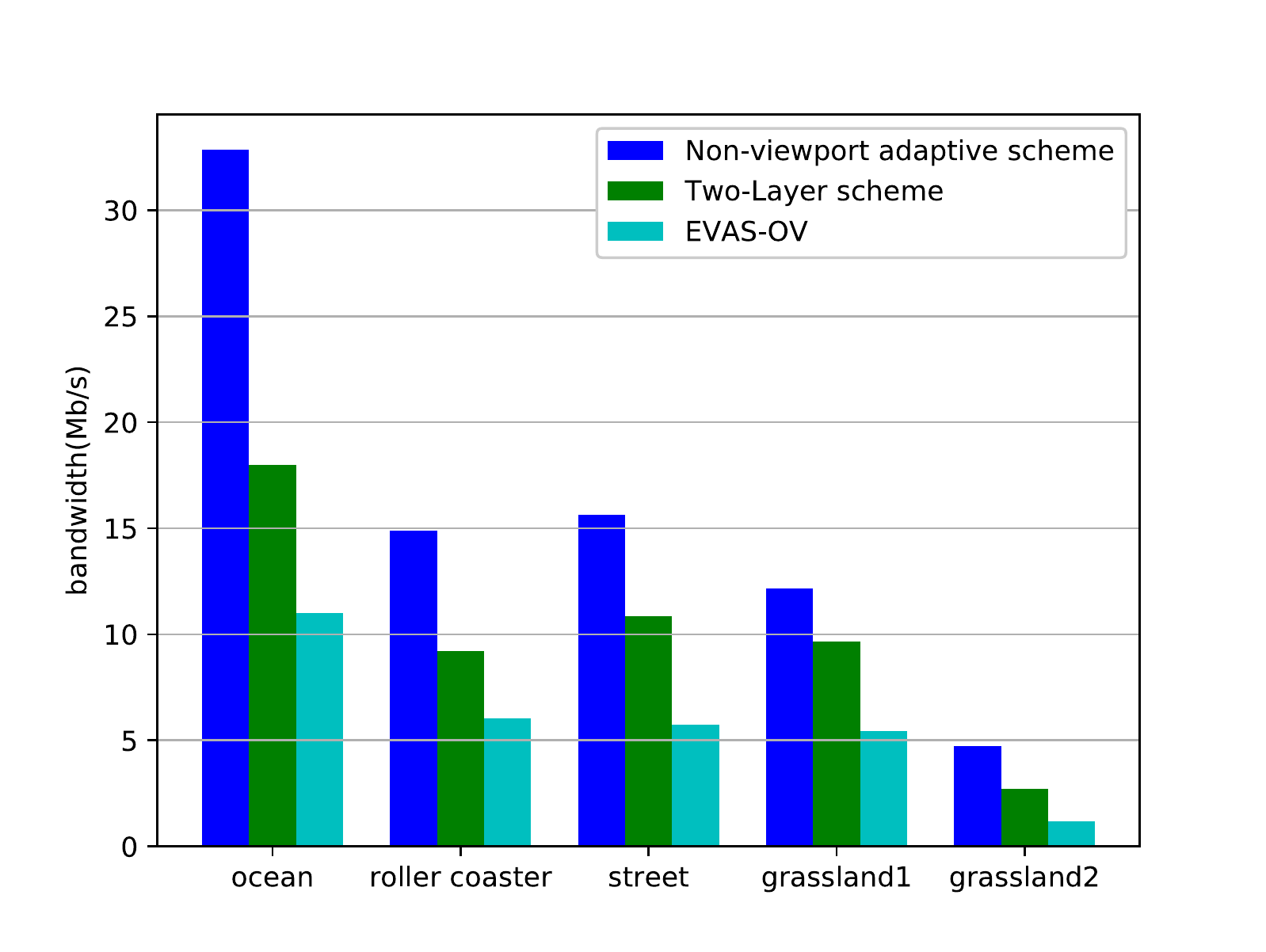}}
	\quad 
	\subfigure[Frame reprojection time]{
		\label{level.sub.2}
		\includegraphics[width=0.45\linewidth]{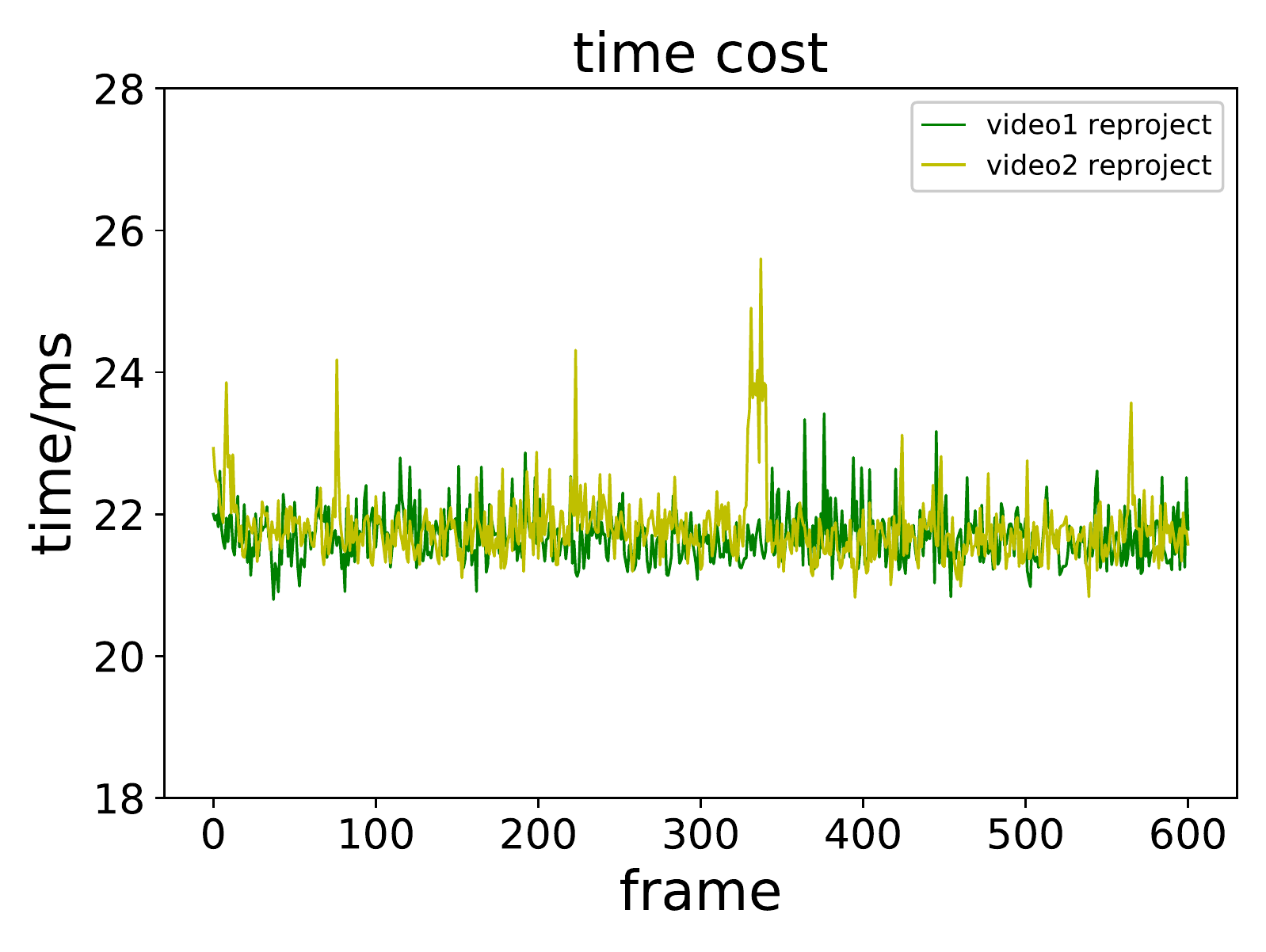}}
	\quad
	\caption{Bandwidth Savings and reprojection time cost}
	\label{level}
\end{figure}

The network bandwidth occupation during transmission is shown in fig 6(a). Compared to the non-viewport adaptive transmission scheme, EVAS-OV saves more than 60\% of bandwidth. Compared with the two-layer transmission scheme, EVAS-OV still achieved a bandwidth savings of about 30\%. It's because we have reduced the size of the base frame. Since the base frame does not appear in a user's FOV area in most cases, its impacts on QoE are small. 

After receiving a VBM frame, the client will reconstruct the VBM frame to generate an OV frame. We use PSNR(Peak Signal to Noise Ratio) and SSIM(Structural Similarity index) to evaluate the quality of the reconstructed OV frame. The green column represents the quality of the user's Fov area after reconstruction, the orange column represents the quality of the margin area and the blue column represents the entire OV frame. As shown in fig 7(a), the FOV area image quality is close to 50dB and PSNR of the reconstruct OV frame is between 25-40dB. Meanwhile, PSNR of margin area is over 30dB, which can ensure the user's QoE when viewports prediction error happened. In fig 7(b), SSIM of the reconstructing OV frame is about 0.92 and SSIM of the FOV area is over 0.98, that's because the FOV frame is cropped directly from the original frame with almost no quality loss.

\begin{figure}[!htbp] 
	\centering  
	\vspace{-0.35cm} 
	\subfigtopskip=5pt 
	\subfigbottomskip=10pt 
	\subfigcapskip=-5pt 

	\subfigure[PSNR of FOV area, Margin area and Reconstruct frame]{
		\label{level.sub.1}
		\includegraphics[width=0.45\linewidth]{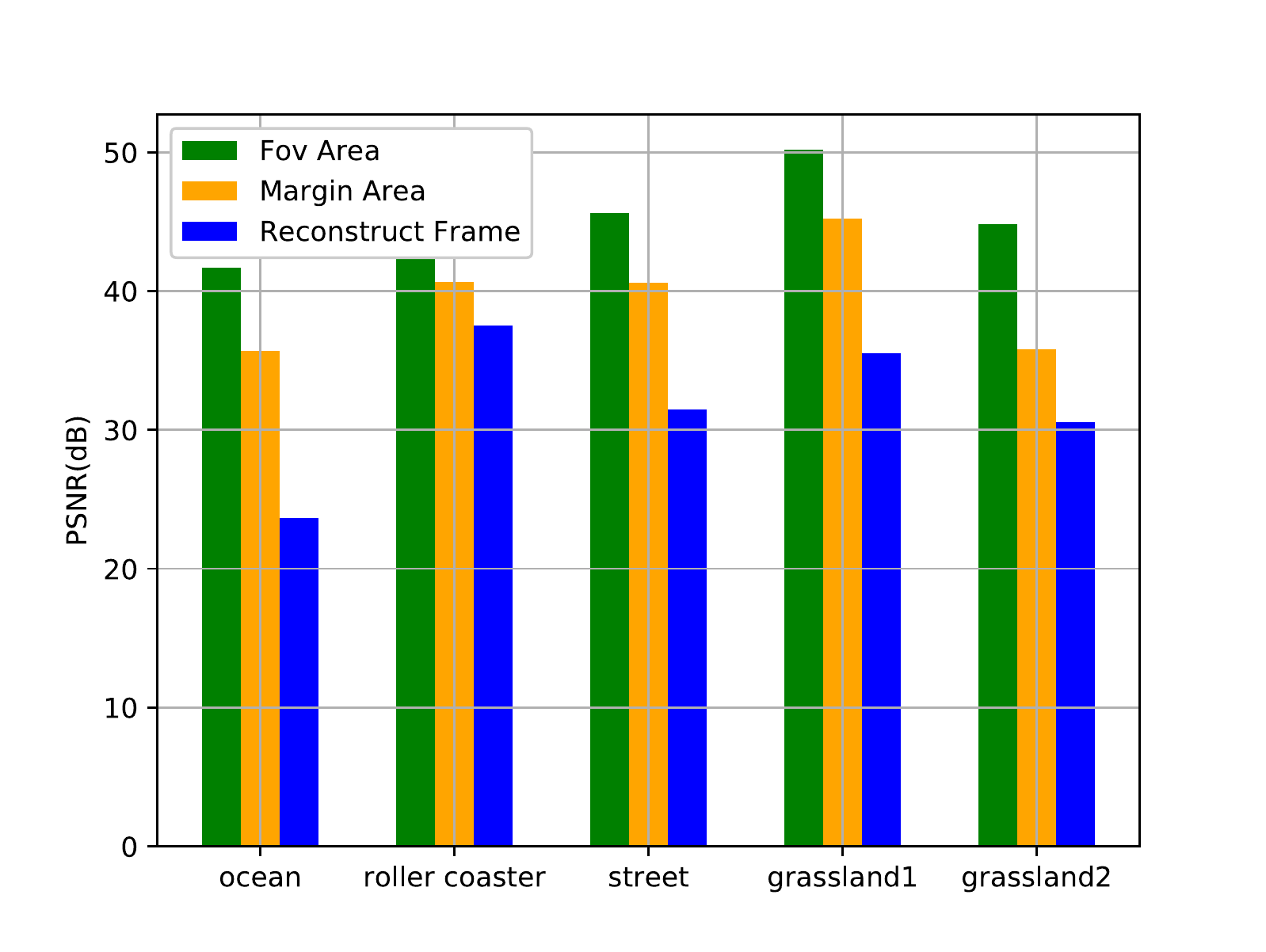}}
	\quad 
	\subfigure[SSIM of FOV area, Margin area and Reconstruct frame]{
		\label{level.sub.2}
		\includegraphics[width=0.45\linewidth]{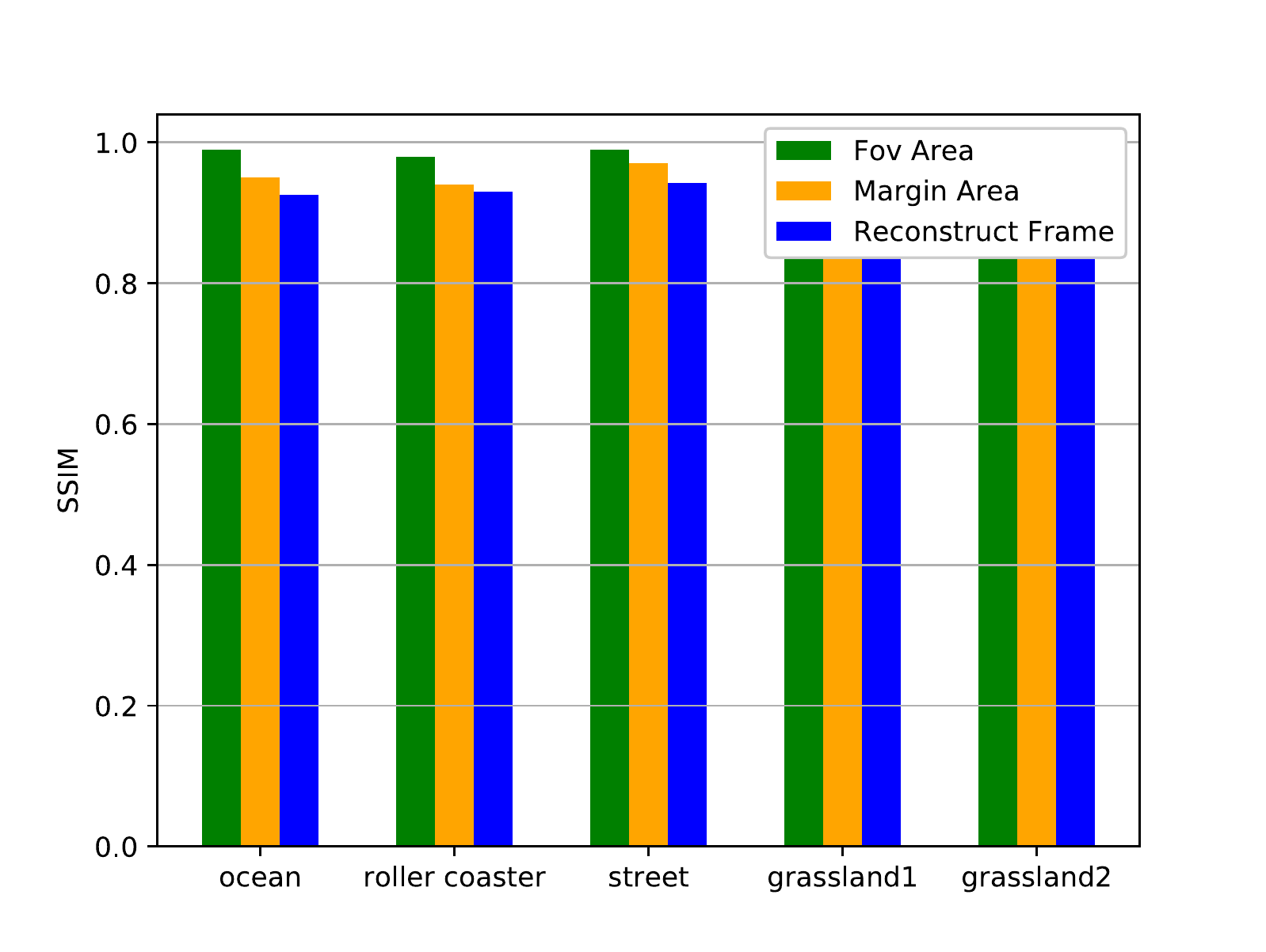}}

 	\caption{PSNR and SSIM}
	\label{level}
\end{figure}

\subsubsection{Computational complexity}
The level of delay determines the experience of OV users. In our scheme, the delay mainly consists of viewport prediction, frame codec, user clustering, frame reprojection, and VBM frame generation. Among them, since reprojection of OV frames needs to be calculated on a matrix of M * N * 3 (M, N is width and height of OV frame), it occupies a major part of the delay. Therefore, we use GPU to speed up this calculation process. We calculate single frame reprojection time of ocean (video1) and street (video2) shown in fig 6(b), and the average processing time is about 22ms.

\section{Conclusion}
In this paper, we propose an edge assisted viewport adaptive transmission scheme EVAS-OV to reduce the bandwidth consumption during OV transmission. To alleviate the computing pressure of the cloud server, computing tasks are offloaded to edge servers. The multicast transmission method after the user clustering further reduces the computing burden of the edge server. By re-projecting the OV frame, we can accurately extract the FOV region from the OV frame, thereby ensuring the consistency of the pixel density in the FOV region. The redundant strategy reduces the impact of viewport prediction errors on users' QoE and improves the robustness of the scheme. We also propose a combined frame strategy VBM frame to facilitate frame encoding and transmission. Experimental results show that compared with the non-viewport adaptive scheme, EVAS-OV saves about 63\% of bandwidth while it doesn't affect the quality of the user's FOV area significantly.

\bibliographystyle{IEEEtran}

\begin{thebibliography}{10}
\providecommand{\url}[1]{#1}
\csname url@samestyle\endcsname
\providecommand{\newblock}{\relax}
\providecommand{\bibinfo}[2]{#2}
\providecommand{\BIBentrySTDinterwordspacing}{\spaceskip=0pt\relax}
\providecommand{\BIBentryALTinterwordstretchfactor}{4}
\providecommand{\BIBentryALTinterwordspacing}{\spaceskip=\fontdimen2\font plus
\BIBentryALTinterwordstretchfactor\fontdimen3\font minus
  \fontdimen4\font\relax}
\providecommand{\BIBforeignlanguage}[2]{{%
\expandafter\ifx\csname l@#1\endcsname\relax
\typeout{** WARNING: IEEEtran.bst: No hyphenation pattern has been}%
\typeout{** loaded for the language `#1'. Using the pattern for}%
\typeout{** the default language instead.}%
\else
\language=\csname l@#1\endcsname
\fi
#2}}
\providecommand{\BIBdecl}{\relax}
\BIBdecl

\bibitem{hormaza2019line}
L.~A. Hormaza, W.~M. Mohammed, B.~R. Ferrer, R.~Bejarano, and J.~L.~M. Lastra,
  ``On-line training and monitoring of robot tasks through virtual reality,''
  in \emph{2019 IEEE 17th International Conference on Industrial Informatics
  (INDIN)}, vol.~1.\hskip 1em plus 0.5em minus 0.4em\relax IEEE, 2019, pp.
  841--846.

\bibitem{petrangeli2017http}
S.~Petrangeli, V.~Swaminathan, M.~Hosseini, and F.~De~Turck, ``An http/2-based
  adaptive streaming framework for 360 virtual reality videos,'' in
  \emph{Proceedings of the 25th ACM international conference on Multimedia},
  2017, pp. 306--314.

\bibitem{graf2017towards}
M.~Graf, C.~Timmerer, and C.~Mueller, ``Towards bandwidth efficient adaptive
  streaming of omnidirectional video over http: Design, implementation, and
  evaluation,'' in \emph{Proceedings of the 8th ACM on Multimedia Systems
  Conference}, 2017, pp. 261--271.

\bibitem{chung2014empirical}
J.~Chung, C.~Gulcehre, K.~Cho, and Y.~Bengio, ``Empirical evaluation of gated
  recurrent neural networks on sequence modeling,'' \emph{arXiv preprint
  arXiv:1412.3555}, 2014.

\bibitem{bao2016shooting}
Y.~Bao, H.~Wu, T.~Zhang, A.~A. Ramli, and X.~Liu, ``Shooting a moving target:
  Motion-prediction-based transmission for 360-degree videos,'' in \emph{2016
  IEEE International Conference on Big Data (Big Data)}.\hskip 1em plus 0.5em
  minus 0.4em\relax IEEE, 2016, pp. 1161--1170.

\bibitem{aladagli2017predicting}
A.~D. Aladagli, E.~Ekmekcioglu, D.~Jarnikov, and A.~Kondoz, ``Predicting head
  trajectories in 360 virtual reality videos,'' in \emph{2017 International
  Conference on 3D Immersion (IC3D)}.\hskip 1em plus 0.5em minus 0.4em\relax
  IEEE, 2017, pp. 1--6.

\bibitem{sun2019two}
L.~Sun, F.~Duanmu, Y.~Liu, Y.~Wang, Y.~Ye, H.~Shi, and D.~Dai, ``A two-tier
  system for on-demand streaming of 360 degree video over dynamic networks,''
  \emph{IEEE Journal on Emerging and Selected Topics in Circuits and Systems},
  vol.~9, no.~1, pp. 43--57, 2019.

\bibitem{shi2019freedom}
S.~Shi, V.~Gupta, and R.~Jana, ``Freedom: Fast recovery enhanced vr delivery
  over mobile networks,'' in \emph{Proceedings of the 17th Annual International
  Conference on Mobile Systems, Applications, and Services}, 2019, pp.
  130--141.

\bibitem{long2018optimal}
K.~Long, C.~Ye, Y.~Cui, and Z.~Liu, ``Optimal multi-quality multicast for 360
  virtual reality video,'' in \emph{2018 IEEE Global Communications Conference
  (GLOBECOM)}.\hskip 1em plus 0.5em minus 0.4em\relax IEEE, 2018, pp. 1--6.

\bibitem{yuan2017ag}
Y.~Yuan, Z.~Zhang, and D.~Liu, ``Ag-ms: A user grouping scheme for dash
  multicast over wireless networks,'' in \emph{2017 IEEE 85th Vehicular
  Technology Conference (VTC Spring)}.\hskip 1em plus 0.5em minus 0.4em\relax
  IEEE, 2017, pp. 1--5.

\bibitem{yang2019cmu}
J.~Yang, J.~Luo, J.~Wang, and S.~Guo, ``Cmu-vp: Cooperative multicast and
  unicast with viewport prediction for vr video streaming in 5g h-cran,''
  \emph{IEEE Access}, vol.~7, pp. 134\,187--134\,197, 2019.

\bibitem{bao2017motion}
Y.~Bao, T.~Zhang, A.~Pande, H.~Wu, and X.~Liu, ``Motion-prediction-based
  multicast for 360-degree video transmissions,'' in \emph{2017 14th Annual
  IEEE International Conference on Sensing, Communication, and Networking
  (SECON)}.\hskip 1em plus 0.5em minus 0.4em\relax IEEE, 2017, pp. 1--9.

\bibitem{kan2019server}
N.~Kan, C.~Liu, J.~Zou, C.~Li, and H.~Xiong, ``A server-side optimized hybrid
  multicast-unicast strategy for multi-user adaptive 360-degree video
  streaming,'' in \emph{2019 IEEE International Conference on Image Processing
  (ICIP)}.\hskip 1em plus 0.5em minus 0.4em\relax IEEE, 2019, pp. 141--145.

\bibitem{corbillon2017360}
X.~Corbillon, F.~De~Simone, and G.~Simon, ``360-degree video head movement
  dataset,'' in \emph{Proceedings of the 8th ACM on Multimedia Systems
  Conference}, 2017, pp. 199--204.

\bibitem{duanmu2017view}
F.~Duanmu, E.~Kurdoglu, Y.~Liu, and Y.~Wang, ``View direction and bandwidth
  adaptive 360 degree video streaming using a two-tier system,'' in \emph{2017
  IEEE International Symposium on Circuits and Systems (ISCAS)}.\hskip 1em plus
  0.5em minus 0.4em\relax IEEE, 2017, pp. 1--4.

\end{thebibliography}

\end{document}